\documentclass{article}

\usepackage{arxiv}

\usepackage{amsmath}
\usepackage{algpseudocode}
\usepackage{algorithm}
\usepackage{multicol}

\usepackage[utf8]{inputenc} 
\usepackage[T1]{fontenc}    
\usepackage{hyperref}       
\usepackage{url}            
\usepackage{booktabs}       
\usepackage{amsfonts}       
\usepackage{nicefrac}       
\usepackage{microtype}      
\usepackage{cleveref}       
\usepackage{graphicx}
\usepackage{natbib}
\usepackage{doi}
\usepackage{enumerate}
\usepackage{tikz} 

\usepackage{multirow}
\usepackage{caption} 
\usepackage{longtable} 
\usepackage{tabularx}
\newcolumntype{L}[1]{>{\raggedright\arraybackslash}p{#1}}
\newcolumntype{C}[1]{>{\centering\arraybackslash}p{#1}}
\newcolumntype{R}[1]{>{\raggedleft\arraybackslash}p{#1}}
\usepackage{relsize} 
\usepackage{arydshln}
\usepackage{adjustbox} 
\usepackage{tabularray}
\usepackage[flushleft]{threeparttable}
\usepackage{xcolor}
\usepackage{subcaption}

\usetikzlibrary{shapes,fit} 
\usetikzlibrary{bayesnet}
\usetikzlibrary{shapes,decorations,arrows,calc,arrows.meta,fit,positioning}
\tikzset{
    -Latex,auto,node distance =1 cm and 1 cm,semithick,
    state/.style ={circle, draw, minimum width = 1.0 cm},
    point/.style = {circle, draw, inner sep=0.04cm,fill,node contents={}}, 
    vmissing/.style={
    draw=none, 
    scale=1,
    text height=0.111cm,
    execute at begin node=\color{black}$\vdots$
    },
    hmissing/.style={
    draw=none, 
    scale=1,
    text height=0.111cm,
    execute at begin node=\color{black}$...$
    },
    bidirected/.style={Latex-Latex,dashed},
    el/.style = {inner sep=2pt, align=left, sloped}
}

\newcommand{\mr}{\mathrm}

\title{
Primed priors for simulation-based validation of Bayesian models
}

\date{}

\newif\ifuniqueAffiliation
\uniqueAffiliationtrue

\ifuniqueAffiliation 
\author{
Luna Fazio\thanks{Shared first authors in alphabetical order.} \\
    Department of Statistics\\
    TU Dortmund University\\
    Germany\\
    \texttt{bmfaziol@gmail.com}
    \And
    Maximilian Scholz$^*$ \\
    Cluster of Excellence SimTech\\
    University of Stuttgart\\
    Germany \\
    \And
    Javier Enrique Aguilar \\
    Department of Statistics\\
    TU Dortmund University\\
    Germany \\
    \And
    Paul-Christian Bürkner \\
    Department of Statistics\\
    TU Dortmund University\\
    Germany\\
}
\else
\usepackage{authblk}

\author[1,3]{%
	{Maximilian Scholz\thanks{\texttt{research.scholz@mailbox.org}}}%
}
\author[2,3]{%
	{Luna Fazio\thanks{\texttt{bmfaziol@gmail.com}}}%
}
\author[2]{
    {Paul-Christian Bürkner\thanks{\texttt{paul.buerkner@gmail.com}}}%
}
 
\affil[1]{Cluster of Excellence SimTech, University of Stuttgart}
\affil[2]{Department of Statistics, TU Dortmund University}
\affil[3]{Shared first authors}
 \fi

\renewcommand{\headeright}{}
\renewcommand{\shorttitle}{Primed priors for Bayesian model validation}

\hypersetup{
pdftitle={Primed priors for Bayesian model validation},
pdfauthor={Luna Fazio, Maximilian Scholz, Javier Enrique Aguilar, and Paul-Christian Bürkner},
pdfkeywords={Bayesian inference, prior specification, data-informed priors, simulation-based calibration},
}

\begin{document}
\maketitle

\begin{abstract}

Simulation-based calibration (SBC) is a method for validating inference algorithms and model implementations through repeated inference on data simulated from a generative model. For a model to be generative, one must specify proper priors. However, in all but the simplest of cases, choosing priors for every model parameter is a nontrivial task. In particular, priors that are too broad can produce numerical issues due to extreme parameter values while overly narrow ones can exclude precisely those regions of the parameter space where legitimate problems in the implementation would have manifested. When the data to be analyzed is already available, the issue can be sidestepped by checking calibration on the corresponding posterior, but that is not always a viable option. In this paper, we adapt the framework of catalytic priors, which have been recently proposed for construction of data-based prior distributions, and propose primed priors, which do not require real data and can therefore facilitate prior specification in SBC. We discuss relevant connections of primed priors to the theory of catalytic priors and show their use for SBC in three simulation studies.
\end{abstract}


\section{Introduction}
\label{sec:introduction}

Presently, users of Bayesian methods enjoy access to a formidable array of modeling frameworks. These provide (i) a probabilistic programming language (PPL) that allows practitioners to specify statistical models with great generality, and (ii) an interface to algorithms for estimating the posteriors implied by those models (see, e.g. Stan, PyMC, Turing; \cite{carpenterStanProbabilisticProgramming2017}, \cite{abril-plaPyMCModernComprehensive2023}, \cite{fjelde2025turing}). As advanced as these tools are, however, even seemingly simple models can entail challenging posteriors which fail to be adequately recovered by the available algorithms. Additionally, human error during model implementation can also be a source of discrepancies between the intended posterior and the one that is actually obtained. Therefore, validation of algorithms and models is a necessary component of a Bayesian workflow \citep{gelman_bayesian_2020}.

Conceptually, simulation-based calibration (SBC) is a straightforward but general procedure for validating an algorithm-model pair \citep{cookValidationSoftwareBayesian2006,talts_validating_2020,modrak_simulationbased_2023}. In SBC, posteriors for a given model are obtained by fitting it to datasets produced by the data-generating process implied by that same model. The self-consistency property of Bayesian models provides a reference against which to compare the resulting posteriors (see details in Section~\ref{sec:sbc_intro}). Discrepancies above a set threshold can then be taken as indicative of a mismatch between the implementations of the inferential and generative model, a posterior that was not adequately recovered by the fitting algorithm, or both.

In practice, however, setting up a data-generating process that works well with SBC is not always easy. On one hand, the priors for the model cannot be too informative, as this will restrict the region of parameter space that is effectively validated with the procedure. On the other hand, priors that are too diffuse will produce large parameter values that lead to computational issues. Anticipating the behavior induced by any particular prior is generally not trivial \citep{mikkola_prior_2023} and doing a search over different hyperparameter specifications becomes impractical as the number of parameters in the model increases.

In parallel work, \cite{sailynojaPosteriorSBCSimulationBased2025} have demonstrated that, if one already has the dataset to be analyzed at hand, a practical way of sidestepping the issue of prior specification in SBC is to first fit the model and then use the resulting posterior for data generation. This provides a check of calibration concentrated exactly on the region of parameter space that is relevant to that analysis. However, this is not always a viable approach: often, the analyst will wish to verify the correct implementation of their model before data arrives, or even as a prerequisite to conduct simulations that will be used to guide data collection. Alternatively, one might be responsible only for model implementation, while the analysis itself is conducted by a separate team; in this case, one would wish to verify model correctness before its delivery, particularly if the end users lack the resources or expertise to conduct validations of their own.

In this paper, we propose a variant of SBC that circumvents the need for an explicit specification of a prior on the model parameters. Briefly, fitting the model to synthetic datasets produces posteriors that can be used as priors with transparently-controlled properties. We denote these \emph{primed priors}, as they are closely related to the framework of catalytic priors \citep{huangCatalyticPriorDistributions2020}, from which we take cues to establish relevant properties and guidelines. In Section~\ref{sec:background}, we provide an overview of the methods that we build on. In Section~\ref{sec:method}, we provide a formal description of primed priors and the modified SBC procedure, along with recommendations for their use. Then, in Section~\ref{sec:case-studies}, we demonstrate the utility of the technique in three case studies based on actual challenges we have previously faced during our statistical practice. Finally, in Section~\ref{sec:discussion}, we summarize the contributions of the paper and highlight related work.

\section{Background}
\label{sec:background}

\subsection{Simulation-based calibration}
\label{sec:sbc_intro}

The core idea of simulation-based calibration (SBC) was proposed in \cite{cookValidationSoftwareBayesian2006}. The aim of SBC is to test the calibration of posterior approximations by taking advantage of the equivalent representations that are admitted by the joint distribution of data $y$ and model parameters $\theta$:
\begin{align}
\label{eq:forward_joint}
  p(y,\theta) &= p(y\mid\theta)p(\theta)\\
\label{eq:inverse_joint}
  &=p(\theta\mid y)p(y).
\end{align}
Following Equation~\ref{eq:forward_joint}, if a procedure draws a sample $\theta_0$ from a proper prior $p(\theta)$ and then generates data according to $p(y\mid \theta_0)$, repeated calls to it will produce samples from $p(y,\theta)$. This is known as the data-generating process and is generally straightforward to implement. Conversely, Equation~\ref{eq:inverse_joint} shows that drawing data $\tilde y$ from the prior predictive distribution $p(y) = \int p(y\mid\theta)p(\theta)d\theta$ and then sampling from the posterior $p(\theta \mid \tilde y)$ should also recover the joint $p(y,\theta)$. In general, obtaining a posterior requires a computational implementation of the model and the use of an approximation procedure. This is the nontrivial component that one is usually interested in verifying the correctness of.

They key insight from \cite{cookValidationSoftwareBayesian2006} is that sampling $(\tilde{y}, \theta_0)$ from $p(y,\theta)$ implies that the distribution that $\theta_0$ follows is $p(\theta \mid \tilde{y})$. Therefore, the rank of the $\theta_0$ used to generate $\tilde y$ must have a uniform distribution within the posterior draws, since $P(\theta<\theta_0)$ corresponds to the probability integral transform. This self-consistency property can then be used to assess the correctness of the posterior approximation.

\subsection{Testing for uniformity in SBC}
\label{sec:sbc_testing}

\cite{cookValidationSoftwareBayesian2006} used a limiting argument on the number of posterior draws to obtain continuous uniformity, from which they derived a $\chi^2$ statistic that can be used for testing. It has since been noted that such an assumption can produce misleading results with the finite draws that must be used in practice \citep{gelmanCorrectionCookGelman2017}. Hence, we discuss the amended procedure proposed by \cite{talts_validating_2020} and better formalized in \cite{modrak_simulationbased_2023} next.

Let $J$ be the number of datasets $\tilde{y}^{(j)}$ ($j = 1, \ldots, J$) generated from $p(y\mid \theta_0^{(j)})$, where each $\theta_0^{(j)}$ is a draw from $p(\theta)$. Let $p_a(\theta\mid\tilde{y}^{(j)})$ denote an approximation of the posterior distribution. We can draw $S$ samples $\theta^{(j,s)}$ $(s = 1, \ldots, S)$ from that approximate posterior. We define the rank statistic $R^{(j)}$ for each univariate quantity of interest as the number of posterior samples that fall below the corresponding ground-truth value $\theta_0^{(j)}$:
\begin{equation}
\label{eq:ranks}
R^{(j)} := \sum_{s = 1}^S \mathbb{I}\left[\theta^{(j,s)} < \theta_0^{(j)}\right].
\end{equation}
This results in a single rank-value per prior sample $\theta^{(j)}_0$ and a distribution of ranks across all $J$ of them. If the approximate posteriors $p_a(\theta \mid \tilde{y}^{(j)})$ are, in fact, recovering the true posteriors $p(\theta \mid \tilde{y}^{(j)})$, the distribution of the ranks $R^{(j)}$ per parameter is discretely uniform.

\begin{figure}
\centering
\includegraphics[width=1\linewidth]{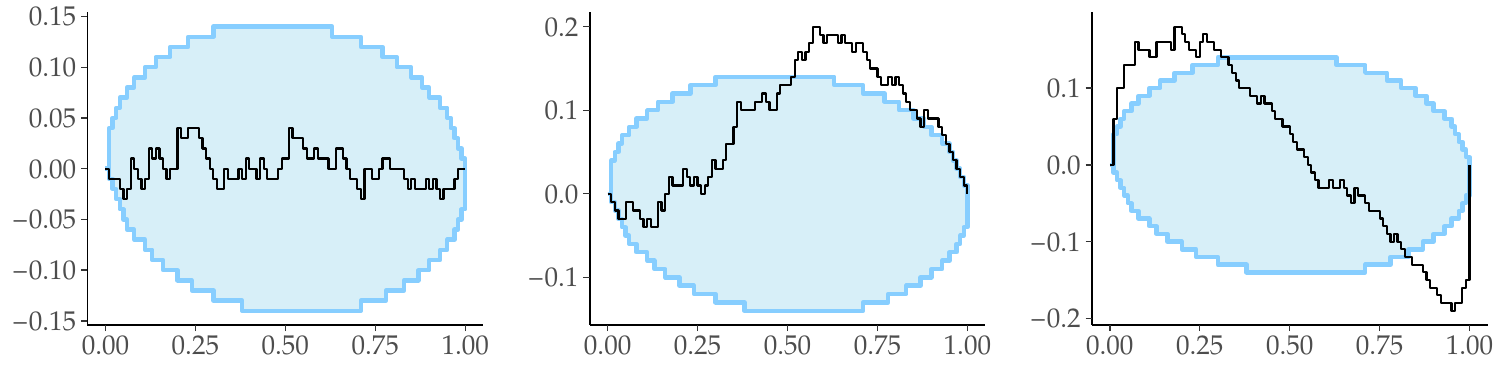}
\caption{Illustrative example of empirical cumulative distribution function (ECDF) difference plots in three calibration scenarios. The blue areas in the ECDF difference plots indicate 95\%-confidence bands under the assumptions of uniformity and thus allow for a null-hypothesis significance test of self-consistent calibration. Left: A well-calibrated quantity. Center: A miscalibrated quantity with too many lower ranks indicating a positive bias in the estimated posteriors. Right: A miscalibrated quantity with too many extreme ranks indicating overconfident posteriors (i.e., variance underestimated).}
\label{fig:sbc-example}
\end{figure}

In order to investigate uniformity in SBC, \cite{talts_validating_2020} proposed using graphical tests. These are constructed by computing simultaneous confidence bands for the empirical cumulative distribution function (ECDF) of the rank distribution under the assumption of uniformity; rejection occurs when the ECDF lies outside the confidence level bands. As ECDF confidence bands can become quite narrow, \cite{sailynoja_graphical_2022} recommended visualizing the ECDF difference instead (see Figure \ref{fig:sbc-example}).

The test statistic implicit in the graphical tests of \cite{talts_validating_2020} was formally characterized in \cite{sailynoja_graphical_2022}. It is denoted as $\gamma$ and corresponds to the probability of observing the most extreme point on the ECDF under the assumption of uniformity:
\begin{equation}
\label{eq:gamma_score}
  \gamma = 2 \underset{j \in \{1, ..., J + 1\}}{\mr{min}} \left( \mr{min}\{\mr{Bin}(R_j \mid J, z_j), 1 - \mr{Bin}(R_j - 1 \mid J, z_j)\} \right),
\end{equation}
where $z_j = \frac{j}{J + 1}$ is the expected proportion of ranks below $j$, $R_j$ is the number of empirical ranks below $j$, and $\mr{Bin}(R \mid J, p)$ is the CDF of the binomial distribution with $J$ trials and a probability of success of $p$, evaluated at $R$. The calculated $\gamma$-score (or its logarithm) can then be compared to a threshold value corresponding to a given confidence level to reject uniformity. We show this later on in the experiments. 

In the absence of uniformity, graphical tests can provide additional insight because the shape of the deviations is related to the way in which the recovered posteriors differ from the target (Figure \ref{fig:sbc-example}). On the other hand, using the $\gamma$ statistic is more practical for models with a large number of parameters, or when comparing calibration across multiple conditions. Hence, the two approaches are complementary.

It should be noted that a fitting procedure which merely returns the prior instead of the correctly updated posterior would still pass SBC as described above. As a way to improve the sensitivity of SBC, \cite{modrak_simulationbased_2023} propose the use of test quantities that can be functions of both multiple parameters and the data. Additionally, they introduce an adjustment that handles ties to enable testing of discrete quantities. The authors provide an R package that implements their recommendations \citep{kimSimulationBasedCalibration}, which we use for our case studies as described in Section~\ref{sec:case-studies}.

We will refer to the procedure, as described up to this point, as standard SBC. We summarize it in Algorithm~\ref{alg:std_sbc}.

\begin{algorithm}
\caption{Standard SBC}\label{alg:std_sbc}
\begin{algorithmic}
\Require $p(y\mid\theta)p(\theta)$
\Comment{Choose a model for which to perform SBC}
\Require $\Call{fit}{}(\cdot)$
\Comment{Choose a procedure to approximate the posterior}
\Require $J > 0$
\Comment{Choose number of SBC replications}
\For{$j \text{ in } 1:J$}
    \State Draw $\theta_0^{(j)}$ from $p(\theta)$
    \Comment{Sample from prior}
    \State Draw $\tilde{y}^{(j)}$ from $p(y \mid \theta_0^{(j)})$
    \Comment{Generate dataset}
    \State $p_a(\theta\mid \tilde y^{(j)}) \gets$ \Call{fit}{$\tilde{y}^{(j)}$}
    \Comment{Obtain posterior approximation}
    \State Draw $\theta^{(j,s)}$ from $p_a(\theta \mid \tilde{y}^{(j)})$ 
    \Comment{Take $S$ samples from approximate posterior}
    \State $R^{(j)} \gets \sum^S_{s = 1} \mathbb{I}[\theta^{(j,s)} < \theta^{(j)}_0]$
    \Comment{Calculate rank statistic}
\EndFor
\State Test $(R^{(1)}, \ldots, R^{(J)})$ for discrete uniformity
\end{algorithmic}
\end{algorithm}

\subsection{Catalytic priors}
\label{sec:catalytic}

We provide a brief overview of catalytic priors here, as the primed priors that we later introduce in Section~\ref{sec:primed_sbc} build on the same core concepts and notation. The goal of catalytic priors, as introduced in \cite{huangCatalyticPriorDistributions2020}, is to "stabilize the estimation of complex {\textquotesingle}working models{\textquotesingle} when sample sizes are too small for standard statistical analysis." The key idea is to augment a given dataset with synthetic data that is generated from the predictive distribution of a simpler model. We reproduce their generic formulation below.

Consider a dataset of size $N$, $y_N$, that is to be analyzed with an observational model $p(y\mid \theta)$, which may be hard to fit. Suppose there is another model $p^*(y \mid \psi)$, with parameters $\psi$, that is in some sense simpler than $p(y\mid \theta)$ and can be reliably fitted using $y_N$. From the implied posterior predictive distribution $p^*(\tilde y^* \mid y_N)$, one can then sample a synthetic dataset of size $M$, $\tilde y^*_M$, and fit the original model to the combined datasets $(y_N, \tilde y^*_M)$. The synthetic data is given a weight of $\tau/M$ where $\tau > 0$ is an additional parameter called prior weight. Its likelihood contribution constitutes the catalytic prior:
\begin{equation}
\label{eq:cat_prior}
  p_{\text{cat}, M}(\theta\mid \tau) \propto p(\tilde y^*_M\mid\theta)^{\tau/M}.
\end{equation}
The intended effect of such a prior is to make fitting of $p(y\mid \theta)$ easier. Clearly, this hinges on actually having a $p^*(y\mid\psi)$ that can be fitted to produce a predictive distribution. The broad principle proposed in \cite{huangCatalyticPriorDistributions2020} for choosing $p^*(y\mid\psi)$ is that its form should be simpler than that of $p(y\mid \theta)$. For instance, as a specific suggestion for regression models, they suggest using intercept-only or dimension-reduced models.

A full specification of the prior requires choosing a size $M$ for the synthetic data and a prior weight $\tau$. \cite{huangCatalyticPriorDistributions2020} provide bounds for the divergence between any given $p_{\text{cat},M}$ and the limiting case when $M \rightarrow \infty$, allowing users to find a compromise between the reduction in finite-sample variation and the increase in computation as $M$ grows. For $\tau$, they provide two strategies: finding a value that minimizes some form of in-sample prediction error, or placing a hyperprior on it. We do not elaborate further on these recommendations because the primed priors that we introduce in Section~\ref{sec:method} have different aims and constraints, even though they involve analogous quantities. We provide our own set of recommendations and discuss where the differences lie in Section~\ref{sec:dataset-size-weight}.

\section{Method}
\label{sec:method}

\subsection{SBC with primed priors}
\label{sec:primed_sbc}

Recall that standard SBC (Algorithm ~\ref{alg:std_sbc}) requires that we generate datasets for $\tilde y$ by sampling from the prior predictive distribution, i.e.
\begin{equation}
\label{eq:y_0}
  \tilde{y} \sim p(y) = \int_\theta p(y\mid\theta) \, p(\theta) \, d\theta.
\end{equation}
We are concerned with situations where using samples from this distribution is not viable. This can happen because the intended specification involves priors that are improper or, despite being proper, place too much probability on extreme parameter values that cause computational issues downstream (usually exhibited when we draw parameters near bounds; see \cite{gabry_visualization_2019,gelman_bayesian_2020}). Such issues may prevent data generation altogether, produce data which does not conform to the model's support, or data that leads to posteriors for which the approximation algorithm badly fails. The latter failure mode could be argued to be among the kinds of discrepancy that SBC aims to detect. That said, one can often be reasonably confident that the algorithm is capable of recovering posteriors in the regions of parameter space that are of practical relevance, and that is the setting that we are interested in here. In that case, failing SBC would not provide useful information about the correctness of our model implementation.

We propose a variant of SBC where, instead of using the original prior directly, one first fits the model of interest $p(y\mid\theta)p(\theta)$ to a synthetic dataset of size $M$, $y'_M$, which receives a weight of $\tau/M$ with $\tau > 0$. The resulting posterior is our \emph{primed prior}:
\begin{equation}
\label{eq:updating}
    p'_M(\theta \mid \tau) \propto p(y_M' \mid \theta)^{\tau/M}\,p(\theta).
\end{equation}
With this new prior, we proceed with SBC in the standard manner: we draw parameter values $\theta'_0 \sim p'_M(\theta\mid\tau)$, we generate datasets $\tilde y' \sim p(y\mid\theta'_0)$, we obtain approximate posteriors $p'_a(\theta\mid\tilde y')$, and then we test the rank statistics for uniformity. For convenient implementation, the target posterior can be obtained by fitting the combined generated and priming datasets on the original model, augmented with the prior weight hyperparameter $\tau$:
\begin{align}
\label{eq:primed_posterior}
    p'(\theta \mid \tilde y'_M, \tau) &\propto
    p(\tilde y' \mid \theta) \, p'_M(\theta\mid\tau)\\ &\propto
    p(\tilde y' \mid \theta) \, p(y_M' \mid \theta)^{\tau/M}\,p(\theta) 
    \label{eq:primed_posterior2}
\end{align}
The description above makes use of a single synthetic dataset $y_M'$ for simplicity. However, in our general formulation of the primed SBC procedure, we allow for a set of $T$ synthetic datasets, $y_M'^{(t)}, t = 1, \ldots, T$, as this offers additional flexibility in the construction of the prior (Section~\ref{sec:dataset-amount}). From each primed prior $p_M'^{(t)}(\theta\mid\tau)$, we can then produce $J^{(t)}$ prior draws $\theta_0'^{(t,j)}$ and SBC datasets $\tilde y'^{(t,j)}$, for a total of $J = \sum_t J^{(t)}$ approximate posteriors and corresponding rank statistics for uniformity testing. We provide recommendations for the origin of the synthetic datasets and the amount of them to use in the following sections. The complete procedure for SBC with primed priors is summarized in Algorithm~\ref{alg:primed_sbc}.

We expect the primed prior to be advantageous as it is usually easier to have a sense of what a reasonable set of observations looks like compared to the set of parameter values which would produce them. Furthermore, the number of observations $M$ and the prior weight $\tau$ can be treated as hyperparameters that offer global control over the informativeness of the updated prior, regardless of the number of parameters in the model.\footnote{The "number of observations" may no longer be a single number in e.g. hierarchical models, but we would generally expect specification of group-level sizes to requires less inputs than those needed for a top-down specification of priors for each model parameter, so some simplicity will be gained regardless.} Introducing the weight $\tau/M$ in the model likelihood requires little modification of the original model implementation, or none at all in the case that the implementation already contemplates the possibility of specifying observation weights.

Our choice of name for primed priors keeps with the chemistry-inspired theme of catalytic priors \citep{huangCatalyticPriorDistributions2020} owing to the similarity of their formulation, and because both are motivated by the goal of enabling a complex model to be fitted. At the same time, it reflects a key difference: instead of introducing a simple model to which existing data is fit, we introduce synthetic data with which the existing model is "primed" (fit). The strategy of regularizing the complex model towards a simpler one used in catalytic priors is one that we also follow, but can only do so through an appropriate choice of synthetic datasets. This is because the choice of model is forced upon us in SBC, since only then are we able to obtain a $p'_M(\theta\mid\tau)$ defined over the set of parameters that needs to be fed forward for generation and testing. We elaborate on choice of values for $M$ and $\tau$ in the following sections.

\begin{algorithm}
\caption{Primed SBC}\label{alg:primed_sbc}
\begin{algorithmic}
\Require $p(y\mid\theta)p(\theta)$
\Comment{Choose a model for which to perform SBC}
\Require $\Call{fit}{}(\cdot)$
\Comment{Choose a procedure to approximate the posterior}
\Require $y_M'^{(t)}$ for $t = 1, \ldots, T$
\Comment{Choose synthetic datasets}
\Require $J^{(t)} > 0$
\Comment{Choose number of SBC simulations per dataset}
\For{$t \text{ in } 1:T$}
\State $p_M'^{(t)}(\theta\mid\tau) \gets$ \Call{fit}{$y_M'^{(t)} \mid \tau$}
\Comment{Obtain primed prior}
\For{$j \text{ in } 1:J^{(t)}$}
    \State Draw $\theta_0'^{(t,j)}$ from $p_M'^{(t)}(\theta\mid\tau)$
    \Comment{Sample from primed prior}
    \State Draw $\tilde y'^{(t,j)}$ from $p(y \mid \theta_0'^{(t,j)})$
    \Comment{Generate dataset}
    \State $p'_a(\theta\mid \tilde y'^{(t,j)}) \gets$ \Call{fit}{$\tilde y'^{(t,j)}, y_M'^{(t)}\mid\tau$}
    \Comment{Obtain posterior approximation}
    \State Draw $\theta^{(t,j,s)}$ from $p'_a(\theta\mid \tilde y'^{(t,j)})$ 
    \Comment{Take $S$ samples from approximate posterior}
    \State $R^{(t,j)} \gets \sum^S_{s = 1} \mathbb{I}[\theta^{(t,j,s)} < \theta_0'^{(t,j)}]$
    \Comment{Calculate rank statistic}
\EndFor
\EndFor
\State Test $(R^{(1,1)}, \ldots, R^{(T,J)})$ for discrete uniformity
\end{algorithmic}
\end{algorithm}

\subsection{Producing synthetic datasets}
\label{sec:dataset-sources}

The properties of the datasets $\tilde{y}'$ simulated from a primed prior $p'_M(\theta \mid \tau)$ are determined by the choice of synthetic data $y'_M$, and that, in turn, determines the region of parameter space that SBC will be effectively testing over. The way in which one produces the synthetic datasets should therefore align with the reason one has for conducting SBC. Accordingly, we differentiate two scenarios based on the availability of real data.

\subsubsection{When real data is available}
\label{sec:sub:dataset-sources-real}

The simplest case is the one in which one already has access to the specific dataset that the model  will be fitted to, and merely wishes to investigate the calibration of the resulting posteriors. In this case, our procedure would not be necessary, and we recommend employing Posterior SBC instead \citep{sailynojaPosteriorSBCSimulationBased2025}.

Now, even if the dataset to be analyzed is not yet available, one can often find data from external sources (e.g. previously conducted studies) that can serve as a rough reference for what a future dataset may look like \citep{schmidli_robust_2014}. In principle, the external data could be directly used in the place of the synthetic data $y'$ that constructs the primed prior. However, if our goal is to test calibration for posteriors that resemble the ones we are likely to obtain once our dataset arrives, we can instead use the external data as a basis for generation of new synthetic datasets that better correspond to the anticipated characteristics of our specific analysis. For instance, if sample size or covariate balance is known ahead of time \citep{linPursuitBalanceOverview2015}, the external data can be stratified and resampled in a way that matches those features. The construction of the primed prior in this case essentially corresponds to that of a power prior \citep{ibrahim_power_2000,ibrahim_power_2015} but the goals are different, as power priors are designed to incorporate historical data in an analysis, while we are interested in assessing calibration of the model.

\subsubsection{When real data is not available}
\label{sec:sub:dataset-sources-fake}

Access to data that covers relevant use cases for the model of interest is not always a given. This is likely to happen, for example, when one wishes to extend some preexisting model. Assessing calibration of the newly developed extension is good practice, but past studies might have only been conducted on data that was deemed appropriate for the models available at the time.

In such a case, one can construct a data-generating process that produces synthetic datasets $y'_M$ with which to obtain the primed prior. Although there are no inherent restrictions on what such a process should be, a sensible starting point is to consider simpler forms of the model to be tested. For example, in a regression model, one may generate from the intercept-only model; in a structural equation model, one may generate from a model that sets the structural coefficients to be zero.\footnote{Defining a formal, generally applicable notion of model complexity is far from trivial. The interested reader may consult \cite{gelman_understanding_2014} and \cite{spiegelhalter_bayesian_2002}.} While applying this heuristic, one should nonetheless remain mindful of the scenarios that the model will be applied to in practice. For instance, our second case study (Section~\ref{sec:case-r2d2}) examines a shrinkage prior in linear regression, and users of such models would certainly be interested in knowing that posteriors remain calibrated in settings where at least some of the coefficients are not zero.

In flexible, highly parameterized models such as polynomial models, generalized additive models, or neural networks, it may not be straightforward to find a parameter configuration that is both simple enough to produce stable fits but complex enough to represent application-relevant features of the data. This is a situation where it makes sense to generate synthetic data from an entirely different model than the one being tested. These flexible models are often used as surrogates of more realistic and interpretable, but hard-to-evaluate models \citep{alizadeh_managing_2020}. For instance, the realistic model might be represented via non-analytic differential equations for which only a simulator is available, making posterior inference a challenge \citep{cranmer_frontier_2020}. However, such a simulator can be used to generate datasets that produce primed priors concentrated in practically relevant regions of the parameter space for our flexible model.

\subsection{Size and weight of synthetic data}
\label{sec:dataset-size-weight}

The size $M$ of a synthetic dataset and the prior weight $\tau$ can be used to control the properties of the resulting primed prior in a straightforward manner, which we consider to be a key practical advantage of the method. We discuss relevant considerations for selecting their values in this section.

The primary requisite for the procedure is the ability to fit the model of interest to the synthetic datasets to recover the primed prior, which must be proper. In the course of discussing priors for model comparison, \cite{berger_intrinsic_1996} observed that datasets of size equal to dimension of the parameter space generally produce proper posteriors if the initial prior was improper. The priors we use in our case studies are all proper, but we found that this heuristic gave good results nonetheless. Therefore, in the absence of some context-specific rationale, we suggest using $\tau = M = \text{dim}(\theta)$ as a starting set of values. If the model still fails to fit, larger values of $M$ can be attempted (still keeping $\tau=M$), though it is sensible at that point to investigate algorithm-specific diagnostics.

Once a value of $M$ that allows a successful model fit has been found, one can assess whether the resulting primed prior covers a satisfactory region of the parameter space \citep{gabry_visualization_2019,gelman_bayesian_2020}. As the total weight $\tau/M$ applied to the synthetic data $y'_M$ scales reciprocally with its size $M$, the informativeness of the data is controlled through the prior weight $\tau$. Under mild assumptions, a larger value of $\tau$ corresponds to a more informative primed prior \citep{ibrahim_power_2015}. 
Therefore, if the prior is found to be too diffuse (concentrated), $\tau$ can be raised (lowered). An intuitive interpretation for $\tau$ is that of the number of observations that the synthetic dataset is effectively equivalent to when updating the initial prior $p(\theta)$. 

Given the above explanation, the use of a prior weight may seem redundant. However, we note that $M$ is discrete, while $\tau$ can take on any positive real value, therefore offering a finer degree of control over the prior. One should also consider that model implementations often involve loops to evaluate likelihood contributions at each observation, so increases in $M$ can slow down the initial fitting process \citep{bardenet2017markov}. Furthermore, for complex data-generating processes, the cost of producing additional observations to increase $M$ may not be trivial (e.g. \cite{gramacy2020surrogates}, Chapter 2).

It is entirely possible that the smallest value of $\tau$ for which the model successfully fits is still found to be too informative. When this is the case, one can use multiple synthetic datasets or fix a subset of parameters during the fitting stage that constructs the primed prior. Both approaches are respectively explained in the next two subsections.

One last conceptual point regarding $M$ has to do with the limiting case $M \rightarrow \infty$. As mentioned in Section~\ref{sec:catalytic}, \cite{huangCatalyticPriorDistributions2020} investigate that case for catalytic priors. They show that (i) under mild assumptions, a unique catalytic prior $p_{\textbf{cat},\infty}$ exists, which they call a \emph{population} catalytic prior, and (ii) the divergence between a given $p_{\textbf{cat},M}$ and $p_{\textbf{cat},\infty}$ is bounded in probability. The same arguments can be followed to show the existence of a unique population primed prior $p'_\infty$. Unlike \citeauthor{huangCatalyticPriorDistributions2020}, however, we do not consider $p'_\infty$ to be of immediate practical relevance and hence do not base any recommendations around this construct. To understand why, one must note that catalytic priors are intended for use in inference; in such a case, it is clear that having inferences depend on the particular set of synthetic data used is not desirable and using large $M$ mitigates this concern. Primed priors, on the other hand, are intended for use in SBC, which is ultimately meant to test discrepancies between a data-generating process and a model implementation. Although the result of the test can certainly depend on the region of the parameter space that is investigated, it seems highly unlikely that the variation across synthetic datasets could result in one primed prior for which calibration is definitively rejected and another one for which it is not (unless, perhaps, one chooses $\tau$ much larger than $M$).

\subsection{Use of multiple synthetic datasets}
\label{sec:dataset-amount}

With a single synthetic dataset, one obtains a single primed prior and this is sufficient to proceed with SBC. Nonetheless, our procedure includes the option to use multiple datasets (Algorithm~\ref{alg:primed_sbc}). The reason to do this is that an increased number of datasets provides better control over the region of parameter space that is effectively tested by the procedure.

For a fixed data-generating process, drawing multiple datasets will simply broaden that region by introducing variation in the primed priors. Alternatively, one may actually wish to reduce variation to increase coverage of particularly relevant regions. For instance, if testing for the presence of an effect is of interest, the probability of covering the case where the effect is absent is zero for any continuous prior on the effect size. Then, it may be desirable to produce a primed prior with synthetic data from a model with a null effect and from one with the effect present.

One could argue that, if broadening the region of parameters to be tested is the goal, reducing the prior weight $\tau$ is a simpler way of doing so. While that is true in principle, one should recall that the context we are interested in is that of models which are not easy to fit. In that case, it may not be possible to lower $\tau$ as much as desired while still obtaining stable fits. Essentially, what we are then doing is to separately explore neighborhoods of the posterior that our algorithms work well within instead of trying to explore a larger region at once. Note that the posterior obtained in this manner is a discrete mixture, with each synthetic dataset producing a corresponding mixture component. Under the workflow we propose in Algorithm~\ref{alg:primed_sbc}, SBC will test calibration on the mixture of posteriors. Although it is possible to test each component individually instead, this introduces issues of sample size and multiple testing; see Section~3.2 of \cite{modrak_simulationbased_2023} for additional discussion.

It is important to note that, because obtaining a primed prior involves fitting the model to the synthetic data, every additional dataset will contribute to the computational cost of the overall procedure. On the other hand, these fits can be run in parallel, so if one is conducting the procedure on, say, a computing cluster, there may not be much of an impact on the overall runtime. Still, there is a tradeoff here to be mindful of.

\subsection{Primed priors for selected parameters}
\label{sec:split-parameters}

So far, we have assumed that the synthetic data $y'_M$ produces a primed prior for all parameters $\theta$. However, in some scenarios, it might be desirable to split $\theta = (\theta_P, \theta_U)$ and only prime a subset of parameters $\theta_P$ while retaining the unprimed (original) prior on the remaining parameters $\theta_U$. For example, in a regression model, we may want to produce a primed prior for all regression coefficients while leaving the intercept's prior unchanged. If the original prior can be factorized into the desired subsets, the partially primed prior takes the form
\begin{equation}
\label{eq:split-prior}
    p'_M(\theta_P,\theta_U\mid\tau) \propto
    p'_M(\theta_P\mid\tau) \, p(\theta_U),
\end{equation}
where $p'_M(\theta_P\mid\tau)$ is the result of marginalizing the full primed prior over the $\theta_U$. Sampling is straightforward: the primed prior is obtained per the original procedure, but only samples for $\theta_P$ are retained, $\theta_U$ is sampled according to its original prior, and one can proceed with generation of datasets $\tilde y'$. An additional modification is needed, however, to restore the correspondence between generating and fitting models.

Recall that the full primed posterior can be obtained by fitting the synthetic and generated data on the original model likelihood (Equation~\ref{eq:primed_posterior2}). If the synthetic data $y'_M$ was used to update only a parameter subset $\theta_P$, then only the generated data $\tilde y'$ should be used to update the subset $\theta_U$ during the fitting step. This can be implemented by introducing a duplicated set of unprimed parameters, $\tilde \theta_U$ and $\theta_U$, so that, when fitting to the combined data $(y'_M,\tilde y')$, the subset $\tilde y'$ is used to update $\tilde \theta_U$ and the subset $y'_M$ is used for a weighted update on $\theta_U$. This leads to the modified likelihood
\begin{equation}
\label{eq:split-likelihood}
p'(\tilde y'_M\mid\theta_P,\tilde\theta_U,\theta_U,\tau) =
    p(\tilde y'\mid\theta_P,\tilde\theta_U) \, p(y'_M\mid\theta_P,\theta_U)^{\tau/M}.
\end{equation}
The corresponding posterior is
\begin{align}
\label{eq:split-posterior}
p'(\theta_P, \tilde\theta_U, \theta_U \mid\tilde y_M',\tau)
    &\propto
p(\tilde y'\mid\theta_P,\tilde\theta_U)\,
p'_M(\theta_P,\theta_U\mid\tau)
p(\tilde\theta_U)\\
    &\propto
p(\tilde y'\mid\theta_P,\tilde\theta_U)\,
p(y'_M\mid\theta_P,\theta_U)^{\tau/M}\,
p(\theta_U)\,p(\theta_P)\,p(\tilde\theta_U).
\end{align}
To test calibration of this partially primed posterior, it suffices to use $\theta_P$ and $\tilde\theta_U$, so draws from $\theta_U$ can be discarded. Since $\theta_U$ plays no role in assessing calibration, one is free to choose different priors for the priming fit and the fit to the generated datasets, i.e. $p(\theta_U) \neq p(\tilde\theta_U)$. This can be particularly helpful when the priming fit itself is difficult, as the posterior may be simplified by setting certain parameters to fixed values. We show an instance of this approach in our third case study (Section~\ref{sec:latent_variable_case}).

\section{Case studies}
\label{sec:case-studies}

In this section, we present three case studies to illustrate the usage of primed priors. All the case studies were written in R \citep{r2024language}, with Stan as the PPL for model specification and its implementation of the No U-Turn Sampler (NUTS) for model fitting \citep{neal_mcmc_2011,homan_nouturn_2014,betancourtIdentifyingOptimalIntegration2016,carpenterStanProbabilisticProgramming2017,standevelopmentteam_stan_2024}.

We rely on the following key packages: rstan and cmdstanr for interfacing with Stan \citep{rstan2024interface, cmdstanr2024}, brms to facilitate Stan code generation \citep{burkner2017brms}, the eponymous SBC package which implements the SBC procedure \citep{modrak_simulationbased_2023}, and bayesim and targets for convenient construction of the simulation pipeline \citep{scholz_bayesim_2024,landauTargets2021}. We additionally use various tidyverse packages for data transformation and plotting \citep{tidyverse2024package}. The complete code can be found in our online appendix~\citep{online_appendix}.

\subsection{SBC for a lower-bounded response}
\label{sec:case_1}

\begin{figure}[!t]
\centering
\includegraphics[width=1\linewidth]{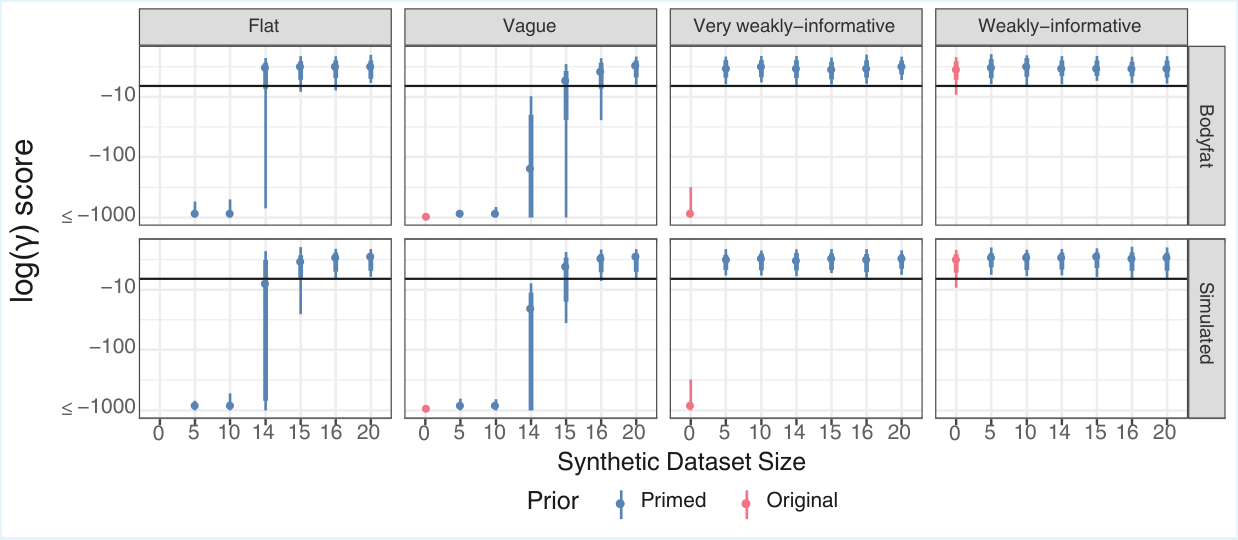}
\caption{Scaling of the $\log(\gamma)$ score of the Gamma regression model described in 
Equation~\eqref{eq:gamma_model} with the different priors described in Equations \eqref{eq:vague}, \eqref{eq:very_weakly_informative} and \eqref{eq:weakly_informative}. Synthetic data was produced by resampling from a reference dataset (bodyfat) or generated from the model itself using a fixed parameter configuration. Plot shows the median as well as two-tailed 66\% and 90\% intervals of the $\log(\gamma)$ scores of all parameters across repetitions. The horizontal line indicates the threshold to reject uniformity at a nominal Type 1 error  rate of 5\% to match the two-tailed 90\% interval. Generated SBC dataset outcomes were censored with a lower threshold of $1e-16$. Standard SBC is not computable for a flat prior in this model as it is improper.}
\label{fig:case_1_scaling}
\end{figure}

Our first case study presents a simple setting, intended to illustrate the basic application of primed priors. We prime a gamma regression model using the bodyfat dataset \citep{johnson_fitting_1996}. After processing, the dataset contains valid observations for 250 men on 13 body measures, which we use as predictors $x_1,\dots,x_{13}$, and an estimate of body fat percentage, which we take as outcome variable $y$. The full model specification is
\begin{equation}
\begin{split}
    \label{eq:gamma_model}
    y_i &\sim \mathrm{Gamma}\left(\alpha, \frac{\alpha}{\mu_i}\right) \quad i = 1,\ldots,250 \\
\log(\mu_i) & = \beta_0 + \sum_{k                         = 1}^{13} \beta_k x_{ki}\\
    \beta_0 & \sim \mathrm{Normal}(2, 5) \\
    \beta_k &\sim \mathrm{Normal}(0, 1) \quad k = 1,\ldots,13 \\
    \alpha & \sim \mathrm{Gamma}(0.1, 0.1), \\
    \end{split}
\end{equation}
where $\alpha$ is a positive shape parameter, the positive rate parameter is $\alpha/\mu$, where $\mu_i$ is the response's mean conditional on observed covariates. 

The computational issue in question arises from the interaction of the weakly informative priors and the log-link on the mean of what should be a strictly positive response. Large negative values of the intercept or coefficients produce response values that, being so close to zero, lead to floating-point underflow such that actual zeros are produced by the data-generating process. Attempting to fit the model on such datasets will immediately produce an error due to the out-of-support observations, so one cannot proceed with standard SBC. As this is a fairly basic regression model, it would not be hard for experienced practitioners to find an alternative specification of priors that remain weakly informative without producing the aforementioned issue, but, for the purpose of illustration, we will look at other options.

Let us discuss two options of simple modifications to the SBC procedure which, in general, could be attempted to sidestep the underflow problem. The first one is to perform rejection sampling, whereby any time a generated dataset $\tilde y^{(j)}$ fails to meet some criterion, the prior sample $\theta_0^{(j)}$ that generated it is redrawn and a new dataset is generated. The process repeats until a dataset that passes the criterion is found. As long as only data-based quantities are used for rejection, calibration will not be affected, but the prior being used will be implicitly modified as the regions likely to produce rejected datasets are being down-weighted (see Section~3.2 of \cite{modrak_simulationbased_2023} and \cite{modrak_rejection_2024}). In our case, we would reject datasets that contain any exact zeros in the response vector. With the priors we are using, however, effectively all generated datasets are rejected, so resampling will not help here.

\begin{figure}[tb]
\centering
\includegraphics[width=1\linewidth]{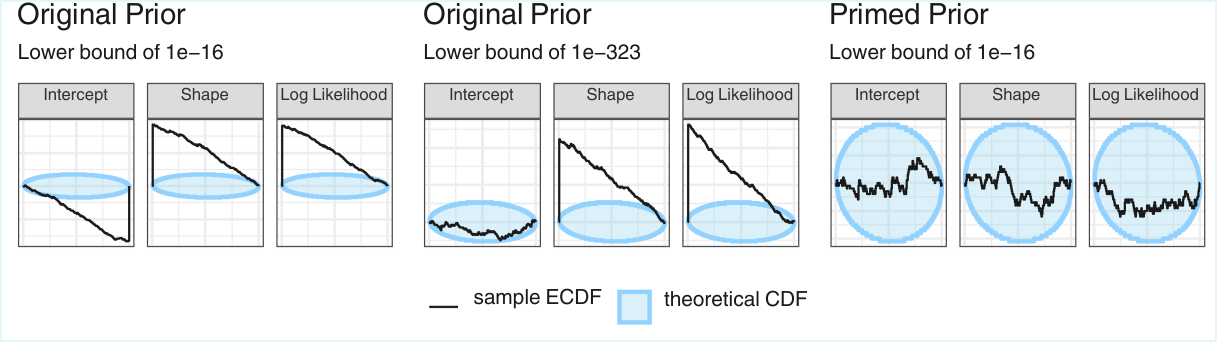}
\caption{ECDF differences of exemplary selected variables for the model described in Equation~\eqref{eq:gamma_model}. The response variable in the generated datasets was left-censored with a threshold of $1e-16$ (left and right) and $1e-323$ (middle). The plots using the original prior on the left and middle display severe calibration problems, while the primed prior ($\tau =M=15$) on the right passes SBC even with censoring.}
\label{fig:case_1_ecdf}
\end{figure}

The other option is to modify the generated datasets by setting values below the rejection threshold to the threshold itself. Doing this introduces a mismatch between the generator and the model so that calibration is formally lost, but the idea is that, if SBC nonetheless passes, we can deem the magnitude of miscalibration to be practically ignorable. However, as shown on Figure~\ref{fig:case_1_ecdf} (left panel) using double-precision ($1e-16$) as the censoring threshold results in failed calibration for all variables. Even using the lowest possible threshold before R rounds numbers to zero (roughly $1e-323$) still results in failed calibration for some model parameters (Figure~\ref{fig:case_1_ecdf}, middle). Accordingly, it would seem that, for this prior specification, it is not practically possible to verify the calibration of the estimated posteriors using standard SBC or simple modifications thereof. Therefore, we apply our primed prior approach.

In order to construct the primed prior, we generate synthetic datasets by simulating from the assumed gamma regression model, using a set of fixed parameter values ($\beta_0 = 1$, $\beta_k = 0.1$ and $\alpha = 1$), and by resampling observations from the bodyfat dataset without replacement (i.e. the "external data" approach for synthetic data generation; Section~\ref{sec:sub:dataset-sources-real}). We find that synthetic datasets of size $\tau=M=15$ (equal to the number of model parameters) are sufficient to produce a primed prior for which all parameters are calibrated, as shown on the right side of Figure~\ref{fig:case_1_ecdf}. This holds even for the higher censoring threshold ($1e-16$), as very small response values become very unlikely under the primed prior and censoring, which would introduce a mismatch with the generating model, no longer occurs in practice.

To show the effect of the primed prior relative to the size of the preconditioning dataset and the width of the original prior,  we ran a small simulation study. The different priors are based on the recommendations detailed in the Stan manual \citep{standevelopmentteam_prior_2023}, namely a completely flat (improper) prior, a vague prior, a very weakly-informative prior, and a standard weakly-informative prior. The latter three are presented in Equations \eqref{eq:vague}, \eqref{eq:very_weakly_informative}, and \eqref{eq:weakly_informative}, respectively. We tested various sample sizes for the synthetic data from $N = 5$ to $N = 30$, with higher resolution around the number of model parame ters (i.e., around $N = 15$).
\begin{multicols}{3}
\noindent
  \begin{equation}
  \begin{split}
  \label{eq:vague}
    \beta_0 & \sim \mathrm{Normal}(2, 100) \\
    \beta_k &\sim \mathrm{Normal}(0, 100) \\
    \alpha & \sim \mathrm{Gamma}(0.01, 0.01)
    \end{split}
  \end{equation}
  \begin{equation}
  \begin{split}
  \label{eq:very_weakly_informative}
    \beta_0 & \sim \mathrm{Normal}(2, 5) \\
    \beta_k &\sim \mathrm{Normal}(0, 1) \\
    \alpha & \sim \mathrm{Gamma}(0.1, 0.1)
    \end{split}
  \end{equation}
  \begin{equation}
  \begin{split}
  \label{eq:weakly_informative}
   \beta_0 & \sim \mathrm{Normal}(2, 1) \\
    \beta_k &\sim \mathrm{Normal}(0, 1) \\
    \alpha & \sim \mathrm{Gamma}(1, 1)
    \end{split}
  \end{equation}
\end{multicols}
We repeated the simulations $T = 10$ times, drawing a new synthetic dataset for each run. We then drew $J = 100$ SBC datasets $\tilde{y}^{(j)}$ with $M=50$ each per run, and drew $S = 1000$ posterior samples per SBC model to calculate the rank statistic $R^{(t,j)}$. The results are shown in Figure~\ref{fig:case_1_scaling}; We calculate the gamma statistic (Equation~\ref{eq:gamma_score} for each model parameter and show their distribution to provide an overview of the results in Figure~\ref{fig:case_1_scaling}. We observe that narrower priors (i.e. larger values of $M$) result in better calibration across all cases. In particular, it is around $M=15$ that calibration noticeably improves even for the flat and vague priors. Full calibration is seen for all $M>20$ (not shown).

Notably, we observed that the model with flat priors achieved better calibration than the model with vague priors, for any fixed preconditioning sample size. This is likely due to the choice of a $\mathrm{Gamma}(0.01, 0.01)$ prior for the shape parameter in the latter case, as it places a significant prior probability on shape values that are close to zero, making the likelihood very challenging numerically. Therefore, it would seem that a primed prior produced from a starting improper priors can produce a stable fit and reach calibration compared to certain proper but very diffuse priors. A related phenomenon is described in \cite{gelman_prior_2006}, where a very wide but simultaneously peaked inverse-gamma prior turns out to be much more informative than intended.

\subsection{SBC for a shrinkage prior}
\label{sec:case-r2d2}

\begin{figure}[t]%
    \centering
    \begin{tikzpicture}

    \definecolor{blauereiterblue}{rgb}{0.0, 0.32, 0.73}    
    \definecolor{blauereitergreen}{rgb}{0.13, 0.55, 0.13}  
    \definecolor{blauereiteryellow}{rgb}{1.0, 0.84, 0.0}   
    \definecolor{blauereiterred}{rgb}{0.86, 0.08, 0.24}    
    \definecolor{blauereitercream}{rgb}{0.98, 0.92, 0.84}  

    \node[state, draw=blauereiterred, fill=blauereiterred!30] (r2) at (0,0) {$R^2$};

    \node[state, draw=blauereiterblue, fill=blauereiterblue!30] (t2) at (2,0) {$\omega^2$};

    \node[state, draw=blauereitergreen, fill=blauereitergreen!30] (phip) at (4,1) {$\phi_1$};
    \node[vmissing] (dots4) at (4,0) {};
    \node[state, draw=blauereitergreen, fill=blauereitergreen!30] (phip1) at (4,-1)  {$\phi_p$};

    \node[state, draw=blauereiteryellow, fill=blauereiteryellow!30] (lambdap) at (6,1) {$\lambda_1^2$};
    \node[vmissing] (dots6) at (6,0) {};
    \node[state, draw=blauereiteryellow, fill=blauereiteryellow!30] (lambdap1) at (6,-1) {$\lambda_{p}^2$};

    \node[state, draw=blauereitercream, fill=blauereitercream!30] (betap) at (8,1) {$\beta_1$};
    \node[vmissing] (dots8) at (8,0) {};
    \node[state, draw=blauereitercream, fill=blauereitercream!30] (betap1) at (8,-1) {$\beta_p$};

    \path (r2) edge (t2);
    \path (t2) edge (phip);
    \path (t2) edge (phip1);
    \path (t2) edge (dots4);

    \path (phip) edge (lambdap);
    \path (phip1) edge (lambdap1);

    \path (lambdap) edge (betap);
    \path (lambdap1) edge (betap1);
\end{tikzpicture}
    \caption{Schematic of the R2D2 prior construction. First, a distribution is assigned to $R^2$, then its uncertainty is propagated to the regression terms via the simplex distribution of the proportions of explained variance. Finally, a distribution is used to allocate variance to each regression term.}
    \label{fig:sr2}
\end{figure}%

In high-dimensional linear regression, continuous global-local (GL) shrinkage priors \citep{vanDP2016Conditions, vanDP2021theoretical} are a class of prior distributions that arise from scale mixtures of normals \citep{west_scale_1987}. These priors admit the hierarchical representation:
\begin{equation}
\label{eqn::glprior}
\beta_i \mid \lambda_i^2, \omega^2, \sigma^2  \sim \mathrm{Normal} \left( 0, \sigma^2 \lambda_i^2 \omega^2 \right), \quad
\lambda_i \sim \pi(\lambda_i), \quad
\omega \sim \pi(\omega), \quad
\sigma \sim \pi(\sigma), \quad
i = 1, \dots, p.
\end{equation}

Here, $\lambda_i$ denotes a local scale parameter specific to each regression coefficient $b_i$, while $\omega$ represents a global scale shared across all coefficients. Priors of the form \eqref{eqn::glprior}, which are centered at zero, promote shrinkage of the regression coefficients towards zero—thus encouraging sparsity. Over the past decade, GL shrinkage priors have become increasingly popular due to both their strong empirical performance and favorable theoretical properties \citep{bayesian_variable_selection_handbook}.

To induce effective shrinkage, the distributions of $\lambda_k$ and $\omega$ are typically chosen so that the marginal distribution of $b_k$ exhibits a sharp peak at zero (to shrink noise) and heavy tails (to retain signals). Although theoretical studies often recommend specific hyperparameter settings to achieve this balance, these choices can introduce practical challenges, particularly regarding computational stability and efficiency.

A prominent example of a shrinkage prior is the R2D2 prior, which places a prior directly on the proportion of variance explained, $R^2$, and allocates the total prior variance $\omega^2 = \frac{R^2}{1 - R^2}$ across coefficients using a Dirichlet distribution \citep{zhang_bayesian_2020}. Assigning a Beta prior to $R^2$ induces a Beta Prime prior on $\omega^2$ with parameters $a_1, a_2 > 0$. The R2D2 prior introduces variance proportions $\phi_i > 0$ for $i = 1, \dots, p$, which follow a Dirichlet distribution with concentration vector $\alpha = (\alpha_1, \dots, \alpha_p)$. These proportions capture the relative importance of the predictors and, together with $\omega^2$, determine the local variances via $\lambda_i^2 =  \sigma^2 \phi_i \omega^2$. While \citet{zhang_bayesian_2020} specify a double-exponential (Laplace) distribution for the coefficients $\beta_i$, we follow the alternative formulation proposed by \citet{aguilar_intuitive_2023}, which assumes normal priors on $\beta_i$. The full hierarchical specification of the R2D2 prior is:
\[
\begin{aligned}
\beta_i \mid \lambda_i^2, \sigma^2 &\sim \mathcal{N}\left(0,\, \sigma^2 \lambda_i^2\right), \\
\lambda_i^2 &=  \sigma^2 \phi_i \omega^2, \\
\phi &\sim \mathrm{Dirichlet}(\alpha), \\
\omega^2 &\sim \mathrm{BetaPrime}(a_1, a_2), \ \ 
\sigma  \sim p(\sigma) 
\end{aligned}
\]
Theoretical results suggest setting $\alpha = (a_\pi, \dots, a_\pi)$ with $a_\pi > 0$ and $a_2 \leq 0.5$ to guarantee strong shrinkage and heavier than Cauchy tails in the marginal priors of the coefficients. \cite{aguilar_intuitive_2023} suggest setting $ a_2 <= 0.5 ,a_\pi = a_1/p$ as a default. The value of $a_1$ can be chosen to set the prior mean and precision of $R^2$.

However, in practice, sampling from a Dirichlet distribution with small concentration parameters is notoriously difficult, often leading to difficulties in sampling and failed SBC diagnostics, particularly as $p$ increases relative to $n$. This gap between theoretical recommendations and practical feasibility presents a fundamental limitation: the hyperparameter values that guarantee optimal shrinkage behavior often render the prior computationally unstable. Consequently, practitioners are typically restricted to using borderline settings (such as $a_\pi = 0.5$ and $a_1= a_2=0.5$) that induce shrinkage while remaining computationally manageable but sacrificing adaptability to $p$ and $n$. Although this compromise allows models to function, it does not fully realize the theoretical advantages envisioned by the original formulation. Ideally, we would like to implement the exact theoretical settings to better align practical applications with the intended prior behavior.

In the following, we demonstrate how primed priors can be used to assess prior calibration in specific regions of the parameter space. We evaluate their behavior across two synthetic scenarios: 1) \textbf{Intercept only model.} We generate data under a model with no covariates, only an intercept. This serves as a baseline scenario where shrinkage dominates and prior influence is minimal. 2) \textbf{Fixed-coefficients model.} Here, we construct coefficient vectors $b$ in which the first 5 and last 5 entries are randomly assigned values from $\{10^2, 10^3, 10^4\}$, with all other entries set to zero. This yields sparse vectors with a small number of large signals. Data are then generated under this sparse linear model.

The objective is to enable the calibration analysis in a setting where sparsity and large signals are present. In such regions, we expect prior characteristics, for instance heavy tails, to have greater influence, in contrast to the intercept only case where the shrinkage effect overwhelms prior structure. This setting is particularly relevant, as shrinkage priors are often evaluated under the assumption that only a small subset of coefficients are nonzero and large in magnitude \citep{castillo_bayesian_2015, van_der_pas_theoretical_2021}. For both models, we fix $n = 100 $, $p = 250$, $M = 100$, $T = 20$, and $J = 5$, and vary the global shrinkage level $\tau$ to examine how prior calibration adapts. We set $a_1$ such that the prior mean of $R^2$ is 0.95 and set $a_2 = 0.5, a_\pi = a_1/p$.

\begin{figure}[!tb]
\centering
\includegraphics[width=1\linewidth]{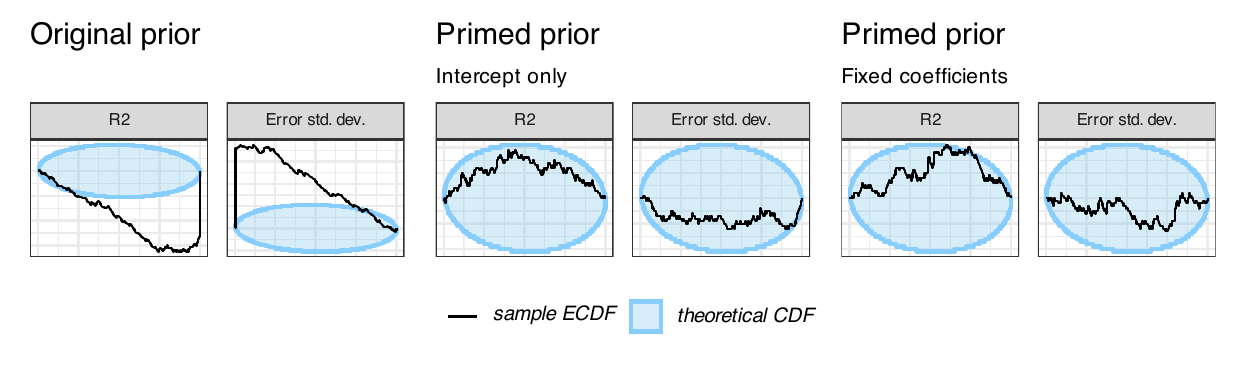}
\caption{ Empirical CDF (ECDF) differences for $R^2$ and $\sigma$ in the shrinkage prior case study. Results are shown for the original prior ($\tau = 0$), which exhibits severe miscalibration, and for both the intercept-only and fixed-coefficients scenarios with $\tau = 0.01$, where calibration improves substantially. Results for $\tau = 0.1$ are similar and omitted for brevity. }
\label{fig:r2d2_plots_1}
\end{figure}

\begin{table}[!h]
\centering
\caption{Posterior summaries of coefficients and local scales under the intercept-only model for varying prior weights $\tau$. In this setting, where all coefficients are zero, the original prior ($\tau = 0$) is severely miscalibrated, but calibration improves even with minimal prior weight ($\tau = 0.01$). As $\tau$ increases, both $\beta$ and $\lambda$ distributions become sharply concentrated around zero, indicating that the prior strongly favors null coefficients.}
\label{tab:r2d2_m1}
\begin{tabular}{ll
  rrr  
  rrr  
  rrr  
}
\toprule
\multicolumn{2}{c}{} &
\multicolumn{9}{c}{Quantiles by Weight $\tau$} \\
\cmidrule(l){3-11}
\multicolumn{2}{c}{} &
\multicolumn{3}{c}{$\tau = 0$} &
\multicolumn{3}{c}{$\tau = 0.01$} &
\multicolumn{3}{c}{$\tau = 0.1$} \\
Variable & Notation & 5\% & 50\% & 95\% & 5\% & 50\% & 95\% & 5\% & 50\% & 95\% \\
\midrule
Error std. dev. & $\sigma$ & 0.1 & 1.1 & 12.6 & 0.5 & 1.2 & 3.9 & 0.7 & 1.0 & 1.5 \\
Local scales & $\lambda$ & 0.0 & 0.5 & 398.6 & 0.0 & 0.0 & 0.0 & 0.0 & 0.0 & 0.0 \\
Explained variance & $R^2$ & 0.6 & 1.0 & 1.0 & 0.1 & 0.4 & 0.8 & 0.0 & 0.2 & 0.5 \\
Coefficients & $\beta$ & -6.4 & 0.0 & 6.9 & -0.0 & 0.0 & 0.0 & -0.0 & 0.0 & 0.0 \\
\bottomrule
\end{tabular}
\end{table}
\begin{table}[htbp]
\centering
\caption{Posterior quantiles of coefficients $\beta$ and local scales $\lambda$ under the fixed-coefficients model for varying prior weights $\tau$. When $\tau = 0$, the prior treats all coefficients similarly, resulting in indistinguishable distributions. At $\tau = 0.01$, the prior begins to differentiate between signal and noise, retaining heavier tails for non-zero coefficients. For $\tau = 0.1$, shrinkage increases for zero coefficients, while the tails for non-zero ones remain heavy, indicating improved calibration in sparse regions of the parameter space.
}
\label{tab:r2d2_m2}
\resizebox{\linewidth}{!}{
\begin{tabular}{ll
  rrr 
  rrr 
  rrr 
}
\toprule
\multicolumn{2}{c}{} &
\multicolumn{9}{c}{Quantiles by Weight $\tau$} \\
\cmidrule(l){3-11}
\multicolumn{2}{c}{} &
\multicolumn{3}{c}{$\tau = 0$} &
\multicolumn{3}{c}{$\tau = 0.01$} &
\multicolumn{3}{c}{$\tau = 0.1$} \\
Variable & Notation & 5\% & 50\% & 95\% & 5\% & 50\% & 95\% & 5\% & 50\% & 95\% \\
\midrule
Error std. dev. & $\sigma$ & 0 & 1.3 & 12.8 & $6.0 \times 10^3$ & $1.52 \times 10^4 $ & $6.77 \times 10^4$ & $1.89 \times 10^3$ & $1.67 \times 10^4$ & $3.08 \times 10^4$ \\
Non-zero scales & $\lambda_{\neq 0}$ & 0 & 0.7 & 395.6 & 0 & 0 & 778.41 & 0 & 0 & 1436.4 \\
Non-zero slopes & $\beta_{\neq 0}$& -5.9 & 0 & 6.1 & -13.09 & 0 & 31.9 & -7.9 & 0 & 323.0 \\
Null scales     & $\lambda_{0} $& 0 & 0.8 & 394.8 & 0 & 0 & 524.4 & 0 & 0 & 193.2 \\
Null slopes     & $\beta_{ 0}$ & -6.1 & 0 & 6.2 & -24.3 & 0 & 14.7 & -6.4 & 0 & 6.8 \\
Explained variance & $R^2$ & 0.6 & 1.0 & 1.0 & 0.04 & 0.42 & 0.8 & 0.01 & 0.21 & 0.94 \\
\bottomrule
\end{tabular}
}
\end{table}

We show the results of using the intercept only model for data generation in Figure~\ref{fig:r2d2_plots_1} and Table \ref{tab:r2d2_m1}. Figure~\ref{fig:r2d2_plots_1} shows that the original prior distribution is severely miscalibrated. This is considerably improved even when using a primed prior with $\tau = 0.01$, i.e. giving the synthetic data a weight equivalent to a single observation. The data we generate focus on the subspace of the parameter space that consists of pure zero coefficient vectors.

Table~\ref{tab:r2d2_m1} further supports this observation. Under the original prior, the distribution of coefficients is centered around zero with mass concentrated from -5 to 5. As soon as $\tau > 0$, the marginal distribution of the coefficients $\beta$ becomes sharply concentrated around zero. This shrinkage is mirrored in the behavior of the local scales $\lambda$ since their distribution shifts toward zero as $\tau$ increases, confirming that all coefficients are essentially null under this prior. This is not a setting we would expect to find in practice, but achieving calibration for this simplified scenario serves as a baseline before moving on to a more elaborate specification.

Figure~\ref{fig:r2d2_plots_1} and Table~\ref{tab:r2d2_m2} show analogous results for the fixed-coefficients model. In this setting, proper calibration is achieved when $\tau = 0.01$. Table~\ref{tab:r2d2_m2} distinguishes between zero and non-zero coefficients and their associated local scales. Here, the focus shifts to a subregion of the parameter space where sparse coefficients include a few strong signals.

For $\tau = 0$, the prior treats all coefficients interchangeably, and the distributions of zero and non-zero coefficients (and their corresponding $\lambda$ values) are nearly identical. At $\tau = 0.01$, the distributions of the non-zero coefficients and their scales retain heavy tails, reflecting the influence of the underlying signal. When the prior weight increases to $\tau = 0.1$, the distribution of local scales for non-zero coefficients exhibits heavier tails, while the scales associated with zero coefficients contract further toward zero. This behavior is also reflected in the coefficient distributions where the tail of the non-zero coefficients becomes heavier, whereas the distribution for zero coefficients becomes more concentrated around zero.

In synthesis, primed priors allowed us to conduct SBC with a shrinkage prior that would have been otherwise hard to work with, while ensuring that relevant regions of the parameter space were covered by the procedure. Overall, we can recommend the following procedure when dealing with a shrinkage prior. First, perform SBC with primed priors constructed from synthetic datasets that match the intended application scenario (e.g., expected sparsity or signal-to-noise ratio). Second, treat the prior weight $\tau$ as a tunable hyperparameter: begin with $\tau = M = \dim(\theta)$, and assess calibration through ECDF rank plots or the $\gamma$-statistic. If calibration remains poor, explore increasing $M$ or modifying the synthetic data to better reflect critical structures (e.g., presence of strong signals). Finally, if inference is the ultimate goal, consider using the primed prior as a diagnostic tool to suggest stable hyperparameter ranges (e.g., $\alpha$, $a_1$, $a_2$ in R2D2).

\subsection{SBC for a latent variable model}
\label{sec:latent_variable_case}

In this case study, we show the application of primed priors to a model that we developed for the purpose of investigating Bayesian estimation of heteroscedastic structural equation models \cite{fazioGaussianDistributionalStructural2025}.

\begin{figure}[!tb]
\centering
\includegraphics[width=0.5\linewidth]{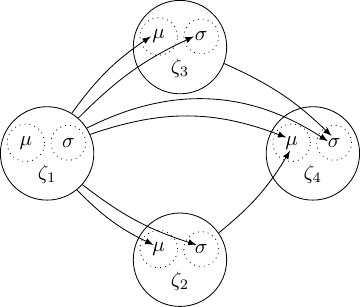}
\caption{Graphical representation of the relationship between the latent variables in our mediation model. $\zeta_1$ has direct and indirect effects on both of $\zeta_4$'s mean and standard deviation parameters through $\zeta_2$ and $\zeta_3$, respectively. These variables are latent and so are only indirectly observed through the measurement model described in Equation~\ref{eq:sem_measurement}.}
\label{fig:sem_diagram}
\end{figure}

Specifically, the model represents a mediation analysis, where the effects that one variable has on the mean and standard deviation of another variable are hypothesized to be potentially explained by changes in intervening variables. We provide a graphical representation of the model in Figure~\ref{fig:sem_diagram} and explain the mathematical details below. Let us use $\zeta_{li}$ to represent the {$i$-th} realization of the {$l$-th} latent variable, with associated parameters $\mu_{li}$ and $\sigma_{li}$. When these parameters are fixed, the second index is omitted. Otherwise, there is an associated linear predictor, with coefficients $\beta_{k\theta_l}$ denoting the slope for the {$k$-th} predictor of the parameter $\theta$ (i.e., $\mu$ or $\sigma$) associated with the {$l$-th} latent variable. The mathematical notation for the latent structural model is
\begin{align}
\begin{split}
\zeta_{1i} &\sim \text{Normal}(0, \sigma_1)\\
\zeta_{2i} &\sim \text{Normal}(\mu_{2i}, \sigma_2)\\
\mu_{2i} &= \beta_{1 \mu_2} \zeta_1\\
\zeta_3 &\sim \text{Normal}(\mu_3, \sigma_3)\\
\mu_3 &= \beta_{1 \mu_3} \zeta_1\\
\zeta_4 &\sim \text{Normal}(\mu_4, \sigma_4)\\
\mu_4 &= \beta_{1 \mu_4} \zeta_1 + \beta_{2 \mu_4} \zeta_2\\
\log \sigma_4 &=
\beta_{0 \sigma_4} + \beta_{1 \sigma_4} \zeta_1 + \beta_{2 \sigma_4} \zeta_3.
\end{split}
\end{align}
We use a Gaussian measurement model, such that the {$j$-th} indicator variable $y_{jli}$ is related to the {$i$-th} realization of the {$l$-th} latent variable through a slope $\lambda_{jl}$ (sometimes called \emph{factor loading}) and an intercept $\nu_{jl}$. The standard deviation of the measurement error is $\tau_{jl}$. The corresponding notation is
\begin{align}
\begin{split}
\label{eq:sem_measurement}
y_{jli} &\sim \text{Normal}(\nu_{jl}+\lambda_{jl} \, \zeta_{li}, \tau_{jl}).
\end{split}
\end{align}
For this simulation, we use $j \in \{1,2,3\}$ indicator variables, $l \in \{1,2,3,4\}$ latent variables, and $i \in \{1,\dots,75\}$ observations per variable. 

\begin{table}
\centering
\caption{Prior specification for the latent variable model and parameters used to generate the synthetic priming datasets.}
    \begin{tabular}{l c c c}
    \hline
    Parameter type & Notation & Weakly informative prior & Parameters for synthetic data \\
    \hline
    Latent mean & &\\
    \quad Slope & $\beta_\mu$ & \text{Normal}(0,\,10) & 0 \\
    \hdashline
    Latent std. dev.& &\\
    \quad Fixed & $\sigma$ & $\text{Gamma}(1,0.5)$ & 1\\
    \quad Intercept & $\beta_{0\sigma}$ & $\text{Exp-Gamma}(1,0.5)$ & 0\\
    \quad Slope & $\beta_{\geq 1\sigma}$ & $\text{Normal}(0,\,2.5)$ & 0\\
    \hdashline
    Indicator parameters& &\\
    \quad Factor loading & $\lambda$ & $\text{Normal}(0,\,10)$ & 1\\
    \quad Intercept & $\nu$ & $\text{Normal}(0,\,32)$ & 0\\
    \quad Error std. dev. & $\tau$ & Gamma(1,\,0.5) & 1\\
    \hline
    \end{tabular}
\label{tab:sempriors}
\end{table}

Now, consider the first set of priors shown in Table~\ref{tab:sempriors}. These are based on the defaults used by blavaan, a package for Bayesian latent variable models \citep{merkleEfficientBayesianStructural2021}. Taken separately, each of the priors constitutes a reasonable weakly informative choice for its corresponding parameter. However, if one examines the implied prior predictive distribution for $\zeta_4$, it turns out to vary over a range that is orders of magnitude larger than those of the preceding variables (Table~\ref{tab:semquantiles}).

\begin{table}
\centering
\caption{Empirical prior predictive distribution of the latent variables using the weakly informative prior. Samples were pooled across the 200 datasets generated by drawing parameters from the prior, each with $N=75$. }
    \begin{tabular}{l|c c c c c}
    \hline
    & \multicolumn{5}{c}{Quantiles} \\
     Variable & 10\% & 25\% & 50\% & 75\% & 90\% \\
    \hline
    $\zeta_1$          & -2.8 & -0.8 & 0 & 0.8 & 2.6 \\
    $\zeta_2, \zeta_3$ & -18.1 & -5.2 & 0 & 5.2 & 18.7 \\
    $\zeta_4$          & $-9.0\times 10^9$ & -178 & -1.4 & 143 & $3.2\times 10^{10}$ \\
    \hline
    \end{tabular}
\label{tab:semquantiles}
\end{table}

This prior specification is already unsatisfactory because it does a poor job of encoding what we believe a reasonable data-generating process would look like, but a bigger problem is that it interacts with the model likelihood to produce a posterior which Stan's MCMC sampler badly fails at recovering. The key challenge is that information about latent variances is only available in the data as a component of the total variance of the indicator variables, which also includes the error~variances~($\tau$):
\begin{align}
\begin{split}
\text{Var}(y_{jli}) = \lambda^2_{jl}\text{Var}(\zeta_{li}) + \tau_{jl}.
\end{split}
\end{align}
The equation above shows that there has to be a direct trade-off between the magnitudes these variances can take. Furthermore, the weakly informative prior implies a high prior probability of both $\sigma_4$ having near-zero values, and of $\sigma_4 < \tau_{j4}$. As a result, the sampler will be stuck in the mode that assigns most of the observed variance to the $\tau_{j4}$ parameters, leading to vastly overestimated $\tau_{j4}$ values and underestimated $\sigma_4$ values (see Table~\ref{tab:semvars}).

\begin{table}
\centering
\caption{We sample from the respective pushforward distributions to recover the implied prior on $\sigma_4$ (first row) and see that, given the Gamma$(1, 0.5)$ prior on $\tau_{j4}$ (second row), this implies $P(\sigma_4 < \tau_{j4}) > 0.5$. The SBC posterior (third row) should recover the prior for $\tau_{j4}$ but instead the sampler favors far larger estimates (convergence also fails badly, so the exact values are not meaningful).}
    \begin{tabular}{r|c c c c c}
    \hline
    & \multicolumn{5}{c}{Quantiles} \\
     Distribution & 10\% & 25\% & 50\% & 75\% & 90\% \\
    \hline
    Implied $p(\sigma_4)$ & $1.8\times 10^{-5}$ & $1.8\times 10^{-2}$ & 1.2 & 67 & $7.5\times 10^4$ \\
    $p(\tau_{j4})$ & 0.2 & 0.6 & 1.4 & 2.8 & 4.6 \\
    $p_{\text{SBC}}(\tau_{j4} \mid y)$ & 113 & $2.3\times 10^7$ & $2.7\times 10^{17}$ & $8.2\times 10^{40}$ & $1.6\times 10^{63}$ \\
    \hline
    \end{tabular}

\label{tab:semvars}
\end{table}

Attempting to perform SBC with such a prior will readily reveal that the resulting posterior approximations are grossly miscalibrated. Rejection sampling will not help here: across all of the simulated datasets, there was not a single converging fit, with Stan's MCMC sampler failing to even initialize in some cases. Once again, there is a clear gap between the models that we can define mathematically and those that we can fit in practice. Fortunately, we have already established that such a weakly informative prior includes parameter combinations that are not of interest to us, so we could try finding a different prior to produce posteriors that are easier to sample from. Reducing the scale of the original priors may seem like a straightforward way of achieving this, but striking the right balance between informativeness and stability would require one to conduct a tedious iterative process for each set of parameters. By contrast, the primed prior approach just requires us to specify a single point in parameter space that we can simulate reasonable data from. From there, one can continue to iterate over the size of the preconditioning sample until the desired criteria are satisfied.

The fixed set of parameter values we used to generate the synthetic data is listed in the last column of Table \ref{tab:sempriors}. We drew $J=200$ SBC datasets $\tilde{y}^{(j)}$, each with $75$ observations and a total of $T=40$ synthetic datasets.

When constructing the preconditioning datasets for this model, we opted to leave in the "true" generated latent variables, i.e. treat them as observed (as discussed in Section~\ref{sec:split-parameters}). This was necessary in order to provide information on the latent scales while keeping the number of preconditioning observations small. At first sight, this may arise concerns regarding the effect on latent variables that will be estimated from the actual data, but one should note that the latent variables are independent from one another and each observation is associated with its own latent variable. Therefore, one can regard the model as already being already factorized in a way that prevents these synthetic values from directly affecting the estimation of the remaining latent variables. Accordingly, it would appear that, for latent variable models, informing only a subset of model parameters is both desirable and easily implemented.

\begin{figure}[!tb]
\centering
\includegraphics[width=1\linewidth]{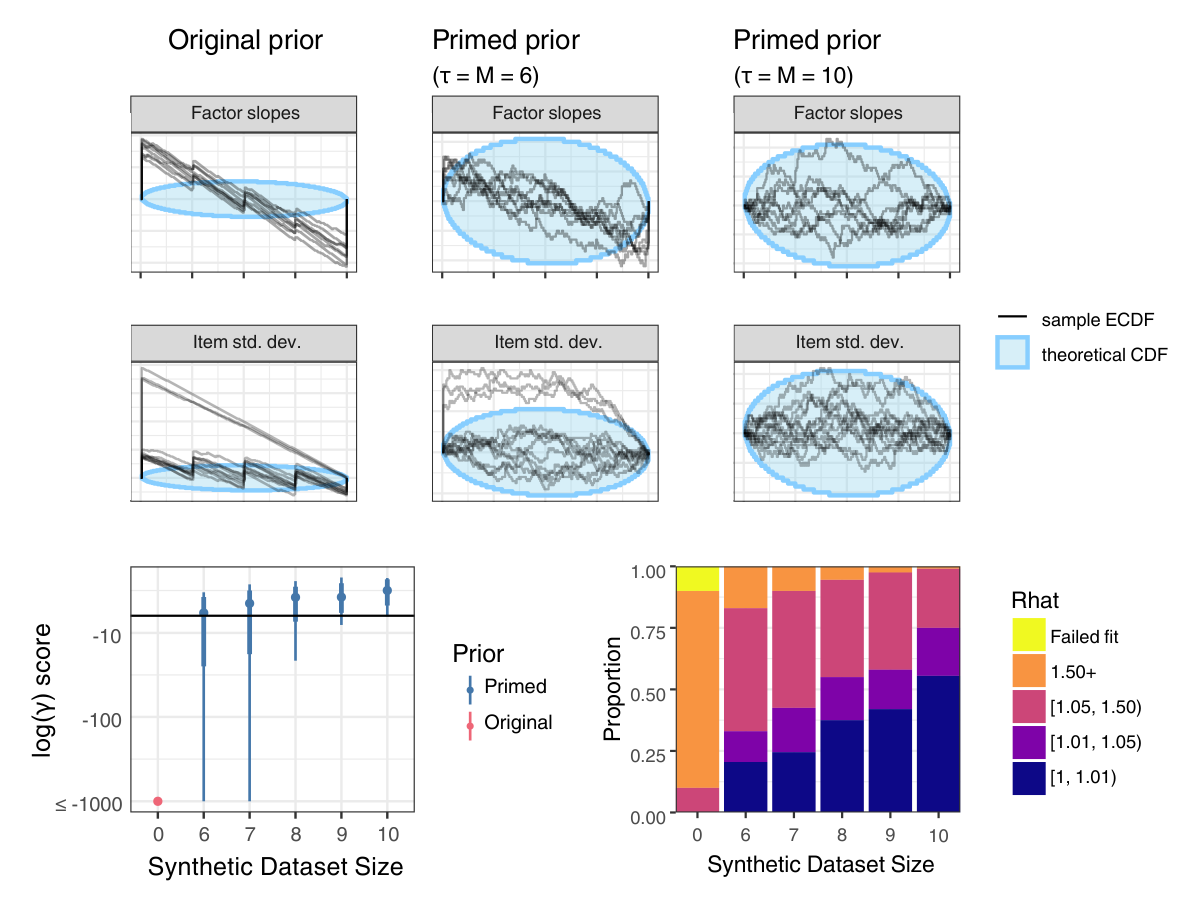}
\caption{Top: ECDF difference plots for selected parameters. The weakly informative prior clearly leads to a miscalibrated posterior. The three tall peaks in the first $\tau$ panel reflect extreme miscalibration of the three $\tau_{j4}$, but we can see problems there propagate to other parameters in the model, including slopes, which are usually of central interest. Notice that the sawtooth pattern with four segments matches the four sampler chains used per fit, which indicates that most fits had a stuck sampler. Bottom left: Plotting the log-gamma statistics for all model parameters allow us to visualize the global improvement in calibration with increasing size of the preconditioning dataset. Bottom right: Convergence also improves with preconditioning.}
\label{fig:sem_plots}
\end{figure}

Our results show that the issues with model miscalibration and non-convergence are promptly resolved with primed priors, even for synthetic datasets of size $M = 10$ (Figure~\ref{fig:sem_plots}). This matches the number of global parameters that characterize the latent variables and the relationships between them, once again matching the heuristic discussed in Section~\ref{sec:dataset-size-weight}.

\section{Conclusion}
\label{sec:discussion}

We have shown that primed priors can facilitate fitting of complex Bayesian models for the purpose of simulation-based model validation, through the use of synthetic data. The set of heuristics given for their construction can significantly ease the task of prior specification by providing a prior weighting parameter $\tau$ which determines the influence of the synthetic data relative to the original prior. Implementation is also straightforward as it only requires concatenating the data being fitted with the observations of the synthetic dataset. 

Primed SBC promotes a targeted, user controlled approach to validation. Standard SBC treats the prior as both a generative object and a source of calibration reference. While this aligns with the Bayesian framework, it limits control over where calibration is assessed, especially in high-dimensional or weakly identifiable models. Primed SBC allows users to specify what kind of parameter configurations are of practical interest, before data is collected, and then directs the SBC procedure to that region.

While we have focused on the utility of primed priors for SBC, this procedure should be relevant for other types of simulation-based checks for Bayesian models such as posterior recovery, prior sensitivity, or sample size estimation. Primed priors may also be helpful for studies focused on model comparison. For example, specifying comparable ground-truths across different data-generating processes can be exceedingly difficult, since prior hyperparameters are hard to match across non-nested model classes \citep{scholz_prediction_2023,scholz_posterior_2023}. By using the same synthetic data to create primed priors for all models, one can avoid artifactual differences that originate from incomparable choices of prior hyperparameters, allowing the focus to remain on the properties of the models of interest. We have investigated a fairly restricted family of models and further investigation into the appropriateness of primed priors for other types of models is warranted. Additionally, it is likely that the heuristics we have provided can be further refined to provide tailored recommendations for specific classes of models and generally obtain a more automated workflow.

\section*{Statements and Declarations}
This work was partially funded by the Deutsche Forschungsgemeinschaft (DFG, German Research Foundation) under Germany’s Excellence Strategy -- EXC-2075 - 390740016 (the Stuttgart Cluster of Excellence SimTech),  DFG Collaborative Research Center 391 (Spatio-Temporal Statistics for the Transition of Energy and Transport) – 520388526, as well as DFG project grants 497785967 and 500663361. The authors gratefully acknowledge the support and funding.


The authors have no competing interests to declare that are relevant to the content of this article.

\section*{Data and code availability}
The data and code that support the findings of this study are openly available on Zenodo at \url{https://zenodo.org/records/13284508}.

\bibliographystyle{unsrtnat}
\bibliography{references}  

\end{document}